%Paper: hep-th/9311058
%From: Melzer Ezer <melzer@ccsg.tau.ac.il>
%Date: Wed, 10 Nov 1993 13:56:14 +0200 (IST)

\input harvmac
\def\ie{{\it i.e.}}
\def\eg{{\it e.g.}}

\def\o{\over}
\def\nl{\hfill\break}
\def\del{\delta}
\def\eps{\epsilon}
\def\th{\theta}
\def\thb{{\th\kern-0.465em \th}}
\def\ZZ{{\bf Z}}

%\draft

\nref\rLuLH{M. L\"uscher, in: {\it Fields, strings, and critical phenomena},
 Les Houches 1988,
 ed. E.~Br\'ezin and J.~Zinn-Justin, (North Holland, Amsterdam, 1989).}
\nref\rKMkink{T.R.~Klassen and E.~Melzer, Nucl.~Phys.~B382 (1992) 441.}
\nref\rTanig{A.B.~Zamolodchikov, Adv.~Stud.~Pure Math.~19 (1989) 1.}
\nref\rModInv{J.L.~Cardy, Nucl.~Phys.~B270 (1986) 186; \nl
 A.~Cappelli, C.~Itzykson and J.-B.~Zuber, Nucl.~Phys.~B280 (1987) 445.}
\nref\rCF{F. Constantinescu and R. Flume, Bonn preprint BONN-HE-93-03.}
\nref\rYZ{V.P.~Yurov and Al.B.~Zamolodchikov, Int.~J.~Mod.~Phys.~A5 (1990)
 3221.}
\nref\rZams{A.B.~Zamolodchikov and Al.B.~Zamolodchikov, Ann.~Phys.~120
 (1980) 253.}
\nref\rLuii{M.~L\"uscher, Commun.~Math.~Phys.~105 (1986) 153;\nl
 M.~L\"uscher and U.~Wolff, Nucl.~Phys.~B339 (1990) 222.}
\nref\rCMS{F.A. Smirnov, Int. J. Mod. Phys. A4 (1989) 4213;  \nl
 J.L.~Cardy and G.~Mussardo, Phys.~Lett.~B225 (1989) 275.}
\nref\rReSm{N.Yu.~Reshetikhin and F.A.~Smirnov, Comm.~Math.~Phys.~131
 (1990) 157.}
\nref\rKedMc{R.~Kedem and B.M.~McCoy,
 J. Stat. Phys.~71 (1993) 865 ~(hep-th/9210129);
 S. Dasmahapatra, R. Kedem, B.M. McCoy and E. Melzer,
 Stony Brook preprint,  hep-th/9304150, J. Stat. Phys. (in press).}
\nref\rKKMMi{R.~Kedem, T.R.~Klassen, B.M.~McCoy and E.~Melzer,
 Phys. Lett. B304 (1993) 263 ~(hep-th/9211102);
 M.~Terhoeven, Bonn preprint, hep-th/9111120;
 A. Kuniba, T. Nakanishi, J. Suzuki, Mod. Phys. Lett. A8 (1993) 1649
 ~(hep-th/9301018);
 R.~Kedem, B.M.~McCoy and E.~Melzer,
 Stony Brook preprint, hep-th/9304056,
 to appear in C.N.~Yang's 70th birthday Festschrift, ed. S.T. Yau.}
\nref\rKKMMii{R.~Kedem, T.R.~Klassen, B.M.~McCoy and E.~Melzer,
 Phys. Lett. B307 (1993) 68  ~(hep-th/9301046).}
\nref\rRR{L.J. Rogers, Proc. London Math. Soc. (ser. 1) 25 (1894) 318;
  (ser. 2) 16 (1917) 315; \nl
 L.J. Rogers and S. Ramanujan, Proc. Camb.
 Phil. Soc. 19 (1919) 211.}
\nref\rSchur{I. Schur, S.-B. Preuss. Akad. Wiss. Phys.-Math. Kl.
 (1917) 302.}
\nref\rYangs{C.N.~Yang and C.P.~Yang, J.~Math.~Phys.~10 (1969) 1115.}
\nref\rZamtba{Al.B.~Zamolodchikov, Nucl.~Phys.~B342 (1990) 695.}
\nref\rKMtba{T.R.~Klassen and E.~Melzer, Nucl.~Phys.~B338 (1990) 485,
 and B350 (1991) 635.}
\nref\rfinsiz{H.W.J. Bl\"ote, J.L. Cardy and M.P. Nightingale, Phys. Rev.
 Lett. 56 (1986) 742;
 I. Affleck, Phys. Rev. Lett. 56 (1986) 746; \nl
 C. Itzykson, H. Saleur and J.-B. Zuber, Europhys. Lett. 2 (1986) 91.}
\nref\rBPZ{A.A.~Belavin, A.M.~Polyakov and A.B.~Zamolodchikov,
  Nucl.~Phys.~B241 (1984) 333.}
\nref\rRocha{A.~Rocha-Caridi, in: {\it Vertex operators in mathematics and
 physics}, ed. J.~Lepowsky et al (Springer, New York, 1985).}
\nref\rFNO{B.L. Feigin, T. Nakanishi and H. Ooguri, Int. J. Mod.
 Phys. A7, Suppl. 1A (1992) 217.}
\nref\rStan{R.P. Stanley, {\it Ordered structures and partitions},
 Mem. Amer. Math. Soc. 119 (1972).}
\nref\rLepWil{J. Lepowsky and R.L. Wilson, Invent. Math. 77 (1984) 199.}
\nref\rBaxHH{R.J. Baxter, J. Stat. Phys. 26 (1981) 427.}
\nref\rLasMar{M.~L\"assig and M.J.~Martins, Nucl.~Phys.~B354 (1991)
 666.}
\nref\rIFT{T.T.~Wu, B.M.~McCoy, C.A.~Tracy and E.~Baruch, Phys.~Rev.~B13
 (1976) 316;~
 B.M. McCoy, C.A. Tracy and T.T. Wu, Phys. Rev. Lett. 38 (1977) 793;~
 M.~Sato, T.~Miwa and M.~Jimbo, Proc.~Japan Acad.~53A
 (1977) 6, 147, 153, 183,  219.}
\nref\rGSO{B. Kaufman, Phys. Rev. 76 (1949) 1232;\nl
 F. Gliozzi, J. Scherk and D. Olive, Nucl. Phys. B122 (1977) 253.}
\nref\rAFZ{A.E. Arinshtein, V.A. Fateev and A.B. Zamolodchikov, Phys. Lett.
  B87 (1979) 389.}
\nref\rGord{B. Gordon, Amer. J. Math. 83 (1961) 363.}
\nref\rAnd{G.E. Andrews, Proc. Nat. Sci. USA 71 (1974) 4082.}
\nref\rFKMS{P.G.O. Freund, T.R. Klassen and E. Melzer,
 Phys. Lett. B229 (1989) 243; \nl
 F.A. Smirnov, Nucl. Phys. B337 (1990) 156.}
\nref\rSKMM{F.A. Smirnov, Int. J. Mod. Phys. A6 (1991) 1407;\nl
 A. Koubek and G. Mussardo, Phys. Lett. B266 (1991) 363.}

\Title{\vbox{\baselineskip12pt\hbox{TAUP 2109-93}\hbox{hep-th/9311058} }}
{\vbox{\centerline{Massive-Conformal Dictionary}}}
\medskip\centerline{Ezer Melzer}
\medskip\centerline{\it School of Physics and Astronomy}
\smallskip\centerline{\it Tel-Aviv University}
\smallskip\centerline{\it Tel-Aviv 69978, ISRAEL}
\medskip\centerline{email: melzer@ccsg.tau.ac.il}
\vskip 13mm

The finite-volume spectrum of an integrable massive perturbation
of a rational conformal field theory interpolates between massive
multi-particle states in infinite volume (IR limit)
and conformal states,
which are approached at zero volume (UV limit).
Each state is labeled in the IR by a set of
`Bethe Ansatz quantum numbers', while in the UV limit it is
characterized primarily
by the conformal dimensions of the conformal field creating it.
We present explicit conjectures for the UV conformal dimensions
corresponding to any IR state
in the $\phi_{1,3}$-perturbed
minimal models ${\cal M}(2,5)$ and ${\cal M}(3,5)$. The conjectures,
which are based on a combinatorial interpretation of the
Rogers-Ramanujan-Schur identities,
are consistent with numerical results obtained previously for low-lying
energy levels.

\Date{\hfill}
\vfill\eject

\newsec{Introduction}
\ftno=0

%{\it 1.} ~
Important properties of a quantum field theory  can be learned from
its spectrum in finite volume.
The volume dependence of energy levels contains information about
the particle content of the theory as well as its $S$-matrix~\rLuLH.
Due to scaling, where the mass scale $M$ times the length $L$
of the system serves
as a dimensionless scaling parameter, one can probe the
UV (massless) limit of the theory by considering the small-volume regime.
In this paper we consider  only  integrable theories in 1+1
dimensions, which have been investigated extensively in recent
years.
A variety of techniques -- perturbative and non-perturbative, analytical
and numerical -- are employed in the study of their finite-volume
spectrum.\foot{Cf.~sect.~1 of~\rKMkink\ for an overview and references.}
(It should be noted, though, that some of the methods used are applicable
to theories in higher dimensions.)

Many interesting integrable theories can be formulated~\rTanig~as
perturbations by a certain relevant spinless operator of
rational conformal field theories (CFTs).
The UV limit in this case is the
CFT itself, which is fairly well understood; in particular, its
finite-volume partition function is given (see \eg~\rModInv) as a
bilinear combination of characters of irreducible highest-weight
representations of the chiral algebra of the CFT.
The spectrum of the perturbed theory can be studied
perturbatively by the
so-called conformal perturbation theory. The corresponding
small-volume expansion is known~\rCF~to
have a finite nonzero radius of convergence (with a finite number,
possibly zero, of terms which
require short-distance regularization). However, due to
the technical difficulties in computing high-order terms, it
is useful only for very small volume,
compared to the inverse mass scale. A numerical non-perturbative
method, known as the truncated conformal space approach~\rYZ, is useful
for getting pretty accurate estimates
up to moderately large volume, but only
for low-lying energy gaps.

On the other hand, if the perturbed CFT is purely
massive the system in infinite volume is described by a relativistic
scattering theory of massive particles~\rTanig~and
the spectrum is built of multi-particle states. In
finite volume the spectrum gets quantized, and
at least at large volume it is still meaningful to talk
about multi-particle states of definite particle content -- they
are stationary scattering states of these particles. If the
theory is integrable, and so its $S$-matrix factorizable~\rZams, then it
is known~\rKMkink\rYZ\rLuii~how to obtain the energies of these states
from the $S$-matrix,
up to off-shell corrections which decay exponentially with the volume.
Unfortunately,
information about these exponential corrections is very limited,
and so the applicabilty of this approximation of the spectrum
is restricted to large volume.
Therefore, it seems that the interesting problem of explicitly
describing the interpolation
between the small and large volume regimes, in other words between the
UV and the IR, remains intractable.

\medskip
Here we make an attempt to
partially solve this problem, namely to  find the
map between conformal states and the multi-particle
states in the massive theory to which they evolve under an
integrable perturbation. Specifically, we will present what we
regard as very plausible conjectures for such
massive-conformal dictionary in two simple -- from
the $S$-matrix point of view -- nontrivial theories,
namely the $\phi_{1,3}$-perturbed nonunitary
minimal models ${\cal M}(2,5)$ (the perturbed Yang-Lee
CFT)~\rCMS~and ${\cal M}(3,5)$~\rReSm.

Our attempt to tackle this problem, which was first addressed
in~\rYZ,\foot{This reference contains many plots which illustrate
the problem most vividly, and the reader is strongly encouraged
to look at it.}
 is motivated by recent developments
in understanding the CFT spectrum from the point of view of the
underlying lattice systems.
The diagonalization of the hamiltonian of certain gapless
spin chains using Bethe equations has been shown~\rKedMc~to
lead in the appropriate scaling
limit to a description of the CFT spectrum in terms of
massless fermionic quasiparticles. This description is encapsulated in
fermionic sum representations~\rKedMc -\rKKMMii,
generalizing the sum side of the
Rogers-Ramanujan-Schur (R-R-S) identities, for the CFT characters.
(In the case of ${\cal M}(2,5)$  the characters are
the two $q$-series of the R-R-S identities themselves~\rRR\rSchur.)
Our strategy is then to
construct a map between the massive multi-particle states and
the conformal multi-quasiparticle states; as we will see,
the  crucial mathematics involved is closely related to Schur's
combinatorial interpretation of the R-R-S identities.

The outline of the paper is as follows. In sect.~2 we review some
general features of the finite-volume spectrum. Sect.~3 discusses in
detail the perturbed ${\cal M}(2,5)$ model, while sect.~4 is devoted
mainly to the perturbed ${\cal M}(3,5)$ model in the phase of
spontaneously broken $\ZZ_2$ symmetry.
Both theories involve a single type of massive particle, hence their
simplicity.
Conclusions and an outlook
to generalizations are included in sect.~5.

\newsec{Finite-volume spectrum (generalities)}

%\medskip {\it 2.} ~
In order to set up a massive-conformal
dictionary we need to recall some general
facts about the finite-volume spectrum, in particular the characterization
of conformal states on the one hand and the massive multi-particle
states on the other. We consider a perturbed CFT on a cylinder
whose circumference $L$ is the ``volume'' of space, and impose
periodic boundary conditions (on some bosonic order parameter) around
the cylinder.
The momentum\foot{Henceforth, the scale $M$ is taken to be the
mass of the lightest particle in the spectrum,
\ie~its rest energy in infinite volume.
Also, state labels, other than 0 which corresponds to the ground state,
will be suppressed for the sake of notational transparency.}
 $P(L; M)$ of any state is therefore some integral multiple
$p$ of $2\pi/L$, and is independent of $M$ since the momentum is
a good quantum number.
For dimensional reasons the energy of an arbitrary state can be written as
\eqn\energy{ E(L;M)~=~{2\pi\o L} e(\rho) ~~~~~~~~~~~~~~~(\rho=ML).}
The dimensionless scaling functions $e(\rho)$ will be referred to
as  scaled energies. We will also introduce the (scaled) energy
gaps with respect to the ground state (the latter is often exactly
calculable from the $S$-matrix using the thermodynamic Bethe
Anstaz~\rYangs -\rKMtba, and is renormalized such that
$e_0(\infty)=0$),
\eqn\gaps{ \hat{E}(L;M)~=~E(L;M)-E_0(L;M)~=~
     {2\pi\o L} \hat{e}(\rho) ~~.}
The  partition function of the theory is defined as
the generating function
\eqn\pf{ Z(\rho)  ~=~ |q|^{e_0(\rho)} \sum_{{\rm states}}
 |q|^{\hat{e}(\rho)}
  \left( {q\o \bar{q}}\right)^{p/2} ~~,}
where $\bar{q}$ is the complex conjugate of $q$.

[Note that the energies, and hence also the gaps, are smooth for positive
volume (the scaled gaps $\hat{e}(\rho)$ are analytic
in $\rho^{2-d_\Phi}$ around zero, where $d_\Phi<2$ is the scaling
dimension of the perturbing field, and their singularities occur
away from the real axis). Therefore it is meaningful to talk
about levels as associated with given functions $\hat{e}(\rho\geq 0)$
despite the fact that in integrable theories many level-crossings
occur at positive $\rho$  even within sectors
of same momentum
(cf.~discussion in~\rYZ).]

Now at the conformal point $\rho=0$, the scaled energy-momentum
of a state which is created by a conformal field of right-left
conformal dimensions $(\Delta,\bar{\Delta})$ is
\eqn\uv{ (\hat{e}(0),~p)~=~ (d-d_{\rm min},~s) ~=~
  (\Delta+\bar{\Delta}-(\Delta+\bar{\Delta})_{\rm min},
   ~\Delta-\bar{\Delta})~,}
where the subscript `min' refers to the field of minimum scaling
dimension which creates the vacuum of the CFT (in a unitary CFT
this is the identity field, $(\Delta,\bar{\Delta})_{\rm min}
=(0,0)$).
Recall~\rfinsiz~that for the ground state ~$e_0(0) = -\tilde{c}/12=
-(c-12d_{\rm min})/12$, where $c$ is the Virasoro central charge
and $\tilde{c}$ is called the effective central charge.
The CFT partition function is the generating function of
conformal dimensions, namely
\eqn\ZCFT{ Z_{\rm CFT} = Z(\rho=0)
 = \sum_{{\rm conf.}\atop{\rm states}}
  q^{\Delta-c/12} ~\bar{q}^{\bar{\Delta}-c/12}
 = |q|^{-\tilde{c}/12} \sum_{{\rm conf.}\atop{\rm states}}
  |q|^{d-d_{{\rm min}}}
   \left( {q\o \bar{q}}\right)^{s/2} .}
It can be expressed as
\eqn\ZCFTch{ Z_{\rm CFT} ~=~ \sum_{i,\bar{i}} N_{i,\bar{i}}
  ~ \chi_i(q) ~\chi_{\bar{i}}(\bar{q})~,}
where the $\chi_i(q)$ are characters of
irreducible highest-weight representations of the chiral algebra of the
CFT, and the $N_{i,\bar{i}}$ are non-negative integers.
Eq.~\ZCFTch\ manifestly shows the decoupling at $\rho=0$
of sectors of right- and left-movers, corresponding to the
$\chi_i(q)$ and $\chi_{{\bar{i}}}(\bar{q})$ respectively.

\medskip
We  now turn to the large-volume regime, and for simplicity consider
the case of a factorizable scattering theory with a single type
of particle (mass $M$), whose vacuum is non-degenerate
in infinite volume (the latter assumption excludes the
possibility of kinks and thus simplifies the analysis, cf.~\rKMkink;
in sect.~4 we will encounter a model with a doubly-degenerate vacuum).
The energy gaps -- as well as the energies
themselves, up to the accuracy stated -- and momenta
of $N$-particle states are given~\rYZ\rLuii~by
\eqn\enmom{ \hat{E}(L;M) ~=~ \sum_{k=1}^N M \cosh\th_k
  + {\cal O}(e^{-\sigma ML})~,
 ~~~~~~~P(L;M) ~=~ \sum_{k=1}^N M \sinh\th_k~~,}
where $\sigma$$>$0 and
the {\it real} rapidities $\th_k$ are quantized via the
following equations of the Bethe Ansatz type:
\eqn\BA{ e^{iML\sinh\th_k} \prod_{{k'=1 \atop k'\neq k}}^N
   S(\th_k-\th_{k'}) ~=~ 1 ~~~~~~~~~~~(k=1,2,\ldots,N),}
where $S(\th)$ is the two-particle scattering amplitude
written as a function of rapidity, as customary~\rZams.
We assume that $S(0)=-1$, which appears
to be universally true in theories where the particle is created
by an interacting bosonic field.\foot{This field statistics, in turn,
is dictated by the periodic boundary conditions we impose, which
correspond to a modular invariant partition function.}
Consequently~\rZamtba, there is an exclusion
rule in rapidity (momentum) space: any solution $\{\th_k\}_{k=1}^N$
of \BA, for any given ~$\rho=ML>0$,
consists of $\th_k$ which are all distinct. This allows us
to order the $\th_k(\rho)$ in any solution such that
{}~$\th_1(\rho)<\th_2(\rho)<\ldots<\th_N(\rho)$.

To analyze eqs.~\enmom-\BA\ it is convenient to write $S(\th)=
-e^{i\tilde{\delta}(\th)}$, where unitarity
{}~$S(\th)S(-\th)=1$~ enables a branch choice in which the
(shifted) phase shift $\tilde{\delta}(\th)$ is an odd function
of $\th$ which is real-valued when $\th$ is real.
Taking the logarithm of eqs.~\BA\ one obtains
\eqn\BAlog{ {\rho\o 2\pi} \sinh\th_k ~=~ n_k - {1\o 2\pi}
    \sum_{{k'=1 \atop k'\neq k}}^N \tilde{\delta}(\th_k-\th_{k'})~,}
with the $n_k$ distinct half-odd-integers (integers) when $N$ is even
(odd).
Since $\tilde{\del}(0)$=0 and $\tilde{\del}(\th)$ is analytic
around 0, it follows that the $\th_k$ are
analytic in $1/\rho$ around 0, and can
be expanded as ~$\th_k=2\pi n_k/\rho+{\cal O}(1/\rho^2)$. Thus,
in particular, the ordering of the $\th_k$ implies  that
the $n_k$ are similarly ordered, $n_1<n_2<\ldots<n_N$. This
argument also suggests that given such a set of $n_k$,
there exists a unique (ordered)
solution $\{\th_k(\rho)\}$ to eqs.~\BAlog\ for
any given $\rho>0$. Such an assumption is customary
in all Bethe-Ansatz-type analyses (for certain $\tilde{\del}(\th)$
it has been actually proven, see \eg~\rYangs), and we will adopt it here.
To summarize, we assume that the $N$-particle
states of the massive theory in finite
volume are unambigously specified by sets of
`Bethe Ansatz quantum numbers'
{}~$\{n_k\}_{k=1}^N\subset \ZZ+{N+1\o 2}$~ with strictly increasing
elements.

Before proceeding to specific models,
let us work out a simple but rather formal
exercise whose result will nevertheless be important later on.
Consider the UV limit $\rho\to 0$ of the
Bethe Ansatz system of eqs.~\enmom, \BAlog\ with
{}~$\tilde{\del}(\th)=\tilde{\del}\cdot{\rm sgn}\th$, where $\tilde{\del}$
is a real constant. One can easily see that in the UV limit
$\th_k\in\{0,\pm\infty\}$, and since the $\th_k$ are ordered such that
{}~$\th_k-\th_{k'}>0$~ for ~$k>k'$, obtain
\eqn\UVlim{\lim_{\rho\to 0} ~{\rho\o 2\pi} \sinh \th_k ~=~
  n_k + (N-2k+1){\tilde{\del}\o 2\pi}~~~~~~~~~~~~
  (k=1,2,\ldots,N).}
The  scaled energy-momentum of the corresponding state
(ignoring any exponential corrections) is
\eqn\UVep{ (\hat{e}(\rho=0),~p) ~=~ \left(\sum_{k=1}^N |p_k|,
 ~ \sum_{k=1}^N p_k\right)~~,}
where ~$p_k = n_k +(N-2k+1){\tilde{\delta}\o 2\pi}$.

\newsec{Perturbed ${\cal M}(2,5)$}

%\medskip  {\it 3.} ~
We now specialize the discussion to the perturbed ${\cal M}(2,5)$ model,
and first summarize some known results for
 the unperturbed CFT. The minimal model~\rBPZ~${\cal M}(2,5)$
is of Virasoro central charge $c=-{22\o 5}$, and
except for the identity field ${\sl 1}$ it
contains only one primary field $\varphi$, whose  conformal dimensions
are $(-{1\o 5}, -{1\o 5})$ so that $\tilde{c}=c-12d_{\varphi}={2\o 5}$.
The partition function is~\rModInv
\eqn\ZYL{ Z_{{\cal M}(2,5)} ~=~ |\chi_0(q)|^2 + |\chi_1(q)|^2
 ~=~ |q|^{-1/30} \left( |\hat{\chi}_0(q)|^2 + |q|^{2/5}
  |\hat{\chi}_1(q)|^2 \right)~,}
where the normalized characters $\hat{\chi}_a$,
with $a$=0 (1) corresponding to $\varphi$ (${\sl 1}$),
are given by~\rRocha\rFNO
\eqn\chihats{ \eqalign{\hat{\chi}_a(q) ~&=~{1\o (q)_\infty}\sum_{k\in\ZZ}
   (q^{k(10k+2a+1)}-q^{(2k+1)(5k-a+2)}) \cr
  &=~ \prod_{n=1}^\infty{1\o (1-q^{5n-4+a})(1-q^{5n-1-a})}
  ~=~ \sum_{m=0}^\infty {q^{m(m+a)} \o (q)_m}~~.\cr} }
Here
\eqn\qsub{ (q)_0~=~1~~,~~~~~~~~(q)_m~=~\prod_{j=1}^m (1-q^j)~~~~~
  {\rm for}~~m=1,2,3,\ldots.}
The equality of the three different-looking $q$-series in \chihats\
constitutes the two R-R-S
identities~\rRR\rSchur.

For us the most physically revealing $q$-series representation for
the two characters $\hat{\chi}_a$
is the last one listed on \chihats, which is
referred to as of fermionic form. The reason is that it has a natural
interpretation in terms of  massless (right-moving, say)
fermionic quasiparticles
occupying certain restricted grids of momentum states, as exhibited
by the following manipulations:
\eqn\qpform{\eqalign{ \sum_{m=0}^\infty {q^{m(m+a)}\o (q)_m} ~&=~
  \sum_{m=0}^\infty ~\sum_{r=0}^\infty Q_m(r)~ q^{r+m(m+2a+1)/2}
   \cr &=~ \sum_{m=0}^\infty
  ~\sum_{ \{I_j\}_{j=1}^m \subset\ZZ_{\geq 0}+ (m+2a+1)/2 \atop
    I_{j+1}-I_j\geq 1} q^{\sum_{j=1}^m I_j}~~,}}
where $Q_m(r)$ is the number of additive partitions of the non-negative
integer $r$ into $m$ distinct non-negative integers, which are
denoted by ~$I_j-{m+2a+1\o 2}$~ on the second line of \qpform\
(see~\rStan~for the identity used in the first line).
A change ~$I_j \mapsto I_j'=I_j-{m+1\o 2}+j$~ in the summation variables
leads to a form which is more suitable for considerations of the
massive perturbation, namely
\eqn\pform{ \hat{\chi}_a(q) ~=~   \sum_{m=0}^\infty
  ~\sum_{ \{I_j'\}_{j=1}^m \subset\ZZ_{\geq a+1} \atop
    I_{j+1}'-I_j'\geq 2} q^{\sum_{j=1}^m I_j'}~~~~~~~~~~~~(a=0,1).}

Note the restriction ~$I_{j+1}'-I_j'\geq 2$~ here,
in contrast to the standard fermionic exclusion rule ~$I_{j+1}-I_j\geq 1$~
in \qpform. This `difference 2 condition'
follows from the `generalized commutation relations' of a so-called
${\cal Z}$-algebra in the Lie theoretic interpretation (and proof)
of the R-R-S identities given in~\rLepWil.
It also plays a prominent role in theorem 3.6 of~\rFNO, which states
that the set
\eqn\basis{ \cup_{m=0}^\infty \left\{
  L_{-I_m'}\ldots L_{-I_1'} v_a~\big|~\{I_j'\}_{j=1}^m \subset
   \ZZ_{\geq a+1},~I_{j+1}'-I_j'\geq 2 \right\} }
forms a basis for the irreducible Virasoro Verma modules of
${\cal M}(2,5)$. (In \basis\ the $L_n$ are the Virasoro generators
and $v_a$, $a=0,1$, are the highest-weight states corresponding to
the primary fields $\varphi$ and ${\sl 1}$, respectively.)
As another side remark, note that a further change of variables
in \pform, to $\{ \sigma_\ell \}_{\ell=1}^\infty$
with $\sigma_\ell$=1 ~if~ $\ell\in \{I_j'\}_{j=1}^m$ and 0 otherwise,
results in the one-dimensional configuration sum representation
for the characters
\eqn\csform{ \hat{\chi}_a(q) ~=~ \sum_{ \{ \sigma_\ell \}_{\ell=1}^\infty,
  ~\sigma_\ell \in \{ 0,1 \} \atop
  \sigma_0=a,~\sigma_{\ell-1}+\sigma_\ell \leq 1}
   q^{\sum_{\ell=1}^\infty \ell\sigma_\ell} ~~,}
which appeared in~\rBaxHH.
To conclude this digression,
let us mention that the equality of the rhs
of \pform\ and the infinite product in \chihats\ is equivalent
to an interesting equality between different restricted partitions of
integers~\rSchur.

\medskip
The importance of eq.~\pform\ for us is revealed when it is
used in \ZYL. Writing first
\eqn\chibar{ \hat{\chi}_a(\bar{q}) ~=~   \sum_{\bar{m}=0}^\infty
  ~\sum_{ \{\bar{I}_j'\}_{j=1}^{\bar{m}} \subset\ZZ_{\leq -(a+1)} \atop
    \bar{I}_{j}'-\bar{I}_{j+1}'\geq 2} \bar{q}^{-\sum_{j=1}^{\bar{m}}
     \bar{I}_j'}~~,}
in order to accomodate the left-mover sector, consider the expression
obtained for $|\hat{\chi}_a(q)|^2$ by multiplying eqs.~\pform\ and
\chibar\ together.
It is a (restricted) sum over two  sets $\{\bar{I}_j'\}$ and
$\{I_j'\}$, which we can combine into a sum over~ $\{p_k\}$, defined
as ~$\{\bar{I}_j'\}\cup \{I_j'\}$~ for $a$=0
and ~$\{\bar{I}_j'\}\cup\{0\}\cup \{I_j'\}$~ for $a$=1.
This leads to
\eqn\ZYLp{ Z_{{\cal M}(2,5)}  ~=~ |q|^{-1/30}
  \sum_{N=0}^\infty ~\sum_{ \{p_k\}_{k=1}^N \subset \ZZ \atop
  p_{k+1}-p_k \geq 2} |q|^{\sum_{k=1}^N (|p_k|+{2\o 5}\del_{p_k,0})}
  \left( {q\o \bar{q}} \right)^{\sum_{k=1}^N p_k/2}~~.}

Eq.~\ZYLp\ suggests an interpretation of
 the full Hilbert space of the CFT ${\cal M}(2,5)$ as built of
multi-particle states characterized by sets of integer momenta
$p_k$ satisfying ~$p_{k+1}-p_k\geq 2$, with a state belonging to the
conformal family $[{\sl 1}]$ or $[\varphi]$ depending on whether
or not, respectively, $0 \in \{p_k\}$.
The scaling dimension and spin of the state corresponding to
$\{p_k\}_{k=1}^N$  are given by
\eqn\dsp{ (d(\{p_k\}),~s(\{p_k\})) ~=~ \left( -{2\o 5}+
  \sum_{k=1}^N \epsilon(p_k),~\sum_{k=1}^N p_k \right)~~,}
where the ``dispersion relation of the conformal particles'' reads
\eqn\eps{ \epsilon(p) ~=~ |p| + {2\o 5} \del_{p,0}~~~~~~~~~~~(p\in\ZZ).}

We now observe that except for the zero-mode energy (the
Kronecker-delta term in \eps), the above description of the spectrum of
${\cal M}(2,5)$ is identical to the content of eqs.~\UVlim-\UVep,
provided we take  $\tilde{\del}=-\pi$ there. To see this, note that
{}~$\{n_k\}_{k=1}^N \in \ZZ+{N+1\o 2}$~ with ~$n_{k+1}-n_k\geq 1$~ implies
{}~$\{p_k\}_{k=1}^N \in \ZZ$~ with ~$p_{k+1}-p_k\geq 2$, if
{}~$p_k = n_k-{N+1\o 2}+k$. Hence we can rewrite \ZYLp\ as
\eqn\ZYLn{ \eqalign{ Z_{{\cal M}(2,5)}  ~=~ & |q|^{-1/30}~
  \sum_{N=0}^\infty ~\sum_{ \{n_k\}_{k=1}^N \subset \ZZ+(N+1)/2 \atop
  n_{k+1}-n_k \geq 1} \cr
  &\times ~|q|^{\sum_{k=1}^N (|n_k-{N+1\o 2}+k|
    +{2\o 5}\del_{n_k-(N+1)/2+k,0})}
   \left( {q\o \bar{q}} \right)^{\sum_{k=1}^N n_k/2}~.\cr} }
By now it is only natural to state a \nl
 {\bf Conjecture}: The $N$-particle state labeled
by ~$\{n_k\}_{k=1}^N \in \ZZ+{N+1\o 2}$ ~($n_1<n_2<\ldots<n_N$)
in the massive perturbation of ${\cal M}(2,5)$, goes over in the
massless limit to a conformal state whose scaling dimension and spin
are
\eqn\dict{ (d,~s)~=~ \left( -{2\o 5}+
  \sum_{k=1}^N \bigl(|n_k-{N+1\o 2}+k|
     +{2\o 5}\del_{n_k-(N+1)/2+k,0}\bigr),
   ~\sum_{k=1}^N n_k \right)~~.}

Following our analysis in reverse
it is possible to further map the $\{n_k\}$
onto the massless quasiparticle labels ~$\{\bar{I}_j\}_{j=1}^{\bar{m}}
\cup \{I_j\}_{j=1}^m$, where the $\bar{I}_j$ and $I_j$ correspond to
the left- and right-moving quasiparticles, respectively, and are
both restricted as in \qpform. In the $[\varphi]$ sector, characterized
by $0\not\in \{p_k=n_k-{N+1\o 2}+k\}$, we have ~$m=N-\bar{m}
=\#\{p_k >0\}$~
and ~$\bar{I}_j=-n_j+{m\o 2}$~ for $j=1,\ldots,\bar{m}$ ~while
{}~$I_j=n_{\bar{m}+j}+{\bar{m}\o 2}$~ for $j=1,\ldots,m$.
In the sector $[{\sl 1}]$, on the other hand, where
{}~$0\in \{p_k=n_k-{N+1\o 2}+k\}$, we have ~$m=N-\bar{m}-1
=\#\{p_k >0\}$~
and ~$\bar{I}_j=-n_j+{m+1\o 2}$~ for $j=1,\ldots,\bar{m}$ ~while
{}~$I_j=n_{\bar{m}+1+j}+{\bar{m}+1\o 2}$~ for $j=1,\ldots,m$.

\medskip
Let us now mention the evidence we have in support of
the conjecture stated above. Introducing the
truncated conformal space approach, Yurov and
Al.~Zamolodchikov~\rYZ~studied numerically the finite-volume spectrum
of the perturbed ${\cal M}(2,5)$ model in the sectors of
total scaled momentum $p$=0,1,2, and compared a total number of
19 low-lying levels in these sectors with the Bethe Ansatz
predictions \enmom-\BA. This allowed them to identify the massive
labels $\{n_k\}$ corresponding to these levels, and our conjecture
is consistent with their findings.

It is interesting to
see the relation between the Bethe Ansatz equations \BA,
which provide a good approximation for the large-volume
spectrum of the perturbed theory, and the UV (zero volume)
quantization condition,
eq.~\UVlim\ with $\tilde{\del}=-\pi$, which was instrumental for
obtaining our conjectured massive-conformal dictionary.
The $S$-matrix of the perturbed ${\cal M}(2,5)$ model,
which should be used in \BA,
is~\rCMS~~$S(\th)={\sinh\th+i\sqrt{3}/2 \o \sinh\th-i\sqrt{3}/2}$~
and the shifted phase-shift appearing in \BAlog\ is
therefore given by ~$\tilde{\del}(\th)=
-2\arctan({2\o \sqrt{3}}\sinh\th)$. The effect of the formal exercise
leading to \UVlim\ can be interpreted as replacing $\tilde{\del}(\th)$
by ~$\tilde{\del}(\infty)\cdot{\rm sgn}\th=-\pi\cdot{\rm sgn}\th$.
We stress that in general this replacement {\it changes} the
result for the $\rho\to0$ limit of the scaled gaps as given
by \enmom-\BA\ (not that {\it a priori} there is any  good reason to trust
these equations at $\rho=0$~!): even though
all ~$\th_k\in\{0,\pm\infty\}$~ in this limit, still differences
between $\th_k$ of the same sign do not diverge, and so the
{}~$\tilde{\del}(\th_k-\th_{k'})$~ in \BAlog\ are not necessarily
evaluated at $\pm\infty$. Note, however, that for one-particle
states, and two-particle states where $\th_1$ and $\th_2$ are not
of the same sign in the
UV limit, the above replacement {\it is} harmless.\foot{It is
also harmless, at least formally, in the thermodynamic Bethe Ansatz
computation of the UV effective central charge, whose value turns out
to depend only on $\tilde{\del}(\infty)$; explicitly~\rZamtba\rKMtba,
$\tilde{c}={6\o \pi^2}{\cal L}({x\o 1+x})$~ where
${\cal L}(z)$ is the Rogers dilogarithm and $x\geq 0$ satisfies
{}~$x=(1+x)^{\tilde{\del}(\infty)/\pi}$.}
 For such states
our conjecture implies that the scaled exponential corrections
to the energy gaps \enmom\ approach exactly 0 or $-d_\varphi={2\o 5}$
in the UV limit. (A similar observation concerning
two-particle states in the zero-momentum sector of several perturbed
rational CFTs was made in~\rLasMar.)

\newsec{Other single-particle models}

%\medskip  {\it 4.} ~
We know of only two other integrable perturbed rational CFTs with
a single type of particle in the spectrum, namely the minimal
models ${\cal M}(3,4)$  and ${\cal M}(3,5)$ perturbed by the
$\phi_{1,3}$ operator. The former is the Ising field theory,
describing the off-critical Ising model (at zero magnetic field)
in the scaling limit~\rIFT.
This theory is rather trivial from the
viewpoint of the spectrum, which can be constructed from that
of a free Majorana fermion by performing the GSO projection~\rGSO.
The full finite-volume spectrum is known exactly. It has been
discussed in detail in~\rKMkink, where the massive-conformal
dictionary in both the high- and low-temperature phases of the theory was
also given. For comparison with eq.~\ZYLn, which  summarizes the
dictionary in the case of perturbed ${\cal M}(2,5)$, let us just
write down the analogous representation of the partition function
of the Ising CFT (from which the dictionary in the high-temperature
phase of the Ising field theory is easily read off):
\eqn\ZIsn{ \eqalign{ Z_{{\cal M}(3,4)} ~=~ & |q|^{-1/24} ~
  \sum_{N=0}^\infty ~\sum_{ \{n_k\}_{k=1}^N \subset \ZZ+(N+1)/2 \atop
  n_{k+1}-n_k \geq 1} \cr
 &\times ~|q|^{(1-(-1)^N)/16+\sum_{k=1}^N |n_k|}
 \left( {q\o \bar{q}} \right)^{\sum_{k=1}^N n_k/2}~.\cr} }
We will therefore  concentrate on ${\cal M}(3,5)$
in the rest of this section.

The perturbed nonunitary model ${\cal M}(3,5)$
has a $\ZZ_2$ symmetry. Like
the Ising field theory it has two phases
(call them the `$\pm$'-phases), depending on the
sign of the coupling to the perturbing field $\phi_{1,3}$,
and the $\ZZ_2$ symmetry is spontaneously broken in one of
them (the `$-$'-phase, say)~\rLasMar.
Unlike the Ising field theory, however, the two phases are not
related by some duality. Little is known about the `$+$'-phase
and we will not discuss it here. The particle spectrum in
the `$-$'-phase consists of a kink and an antikink,
which interpolate between two degenerate vacua.
The amplitude  for scattering of a kink on an antikink
(or vice versa) is
given in~\rReSm~as ~$S(\th) =
-i\tanh({\th\o 2}-{i\pi\o 4})$, so that
{}~$\tilde{\delta}(\th) = \arctan(\sinh\th)$~ with
{}~$\tilde{\delta}(\pm\infty) = \pm{\pi\o 2}$.

In~\rLasMar~the finite-volume spectrum of the theory
was studied using
the truncated conformal space approach. Results for
several low-lying levels in the zero-momentum sector were compared
with predictions of the Bethe Ansatz equations \enmom-\BA\ for
two-kink states. With periodic boundary conditions, which we restrict
attention to, there are only even-$N$-kink states
(${N\o 2}$ kinks and ${N\o 2}$ antikinks) in the spectrum,
and eq.~\BAlog\ should be modified (cf.~\rKMkink) to allow for all sets
{}~$\{n_k\}_{k=1}^{N}\subset {1\o 2}\ZZ$~ with $N$ even and
{}~$n_{k+1}-n_k \in \ZZ_{\geq 1}$. The results
of~\rLasMar~enable identification of the UV scaling dimensions
{}~$d={1\o 5},{3\o 4},1{1\o 5},1{3\o 4},2{1\o 5},2{3\o 4}$ ~as
corresponding to the two-kink states $\{-n,n\}$ with~ $n={1\o 2},1,
{3\o 2},2,{5\o 2},3$, respectively, in our notation.

Now the central charge of ${\cal M}(3,5)$ is $-{3\o 5}$, and
there are four primary fields $\phi_{1,r}$ whose left=right conformal
dimensions and $\ZZ_2$-parities are ~$\Delta^C=
0^+, -{1\o 20}^-, {1\o 5}^+,{3\o 4}^-$~ for ~$r=1,2,3,4$,
respectively. Hence the vacuum
is created by the $\ZZ_2$-odd field $\phi_{1,2}$,
and the effective central charge is ${3\o 5}$.
The CFT partition function is
\eqn\ZYL{ Z_{{\cal M}(3,5)} ~=~ \sum_{r=1}^4|\chi_{1,r}(q)|^2
 ~=~ |q|^{-1/20} \left( \sum_{\ell=0}^1 |\hat{\chi}^{(\ell)}_0(q)|^2
    + |q|^{1/10} \sum_{\ell=0}^1 |\hat{\chi}^{(\ell)}_1(q)|^2\right)~,}
where (see~\rKKMMii~and references therein)
\eqn\chars{ \hat{\chi}_a^{(\ell)}(q) ~=~ \sum_{m=0 \atop
   m\equiv \ell({\rm mod}~2)}^\infty {q^{m(m+2a)/4} \o (q)_m}
  ~~~~~~~~~~~(a,\ell=0,1).}
The labels ~$(a,\ell)$=(0,0),~(0,1),~(1,0),~(1,1) ~
correspond to ~$r$=2, 3, 1, 4.
%; the $\ZZ_2$ charge is therefore given by $C=(-)^{a+\ell+1}$.
As in the case of ${\cal M}(2,5)$, the fermionic quasiparticle
representation \chars\ for the characters of ${\cal M}(3,5)$
is our starting point for obtaining
the conjectured massive-conformal dictionary.

Similarly to \qpform-\pform, we first write
\eqn\manip{ \eqalign{ {q^{m(m+2a)/4} \o (q)_m} ~&=~
 \sum_{ \{I_j\}_{j=1}^m \subset
 \ZZ_{\geq 0} -(m-2a-2)/4 \atop I_{j+1}-I_j \geq 1} q^{\sum_{j=1}^m I_j}\cr
 ~&=~ \sum_{ \{I_j'\}_{j=1}^m,~ I_j' \in \ZZ_{\geq 0}-(2j-2a-3)/4
    \atop I_{j+1}'-I_j' \geq 1/2} q^{\sum_{j=1}^m I_j'}~~,\cr} }
where the change of variables ~$I_j \mapsto I_j'=I_j+{m+1\o 4}-{j\o 2}$~
has been performed. Putting together the sectors of left- and
right-movers we arrive at
\eqn\Zthfi{ Z_{{\cal M}(3,5)} =
 |q|^{-{1\o 20}} \sum_{N=0 \atop N~{\rm even}}^\infty
 \sum_{ \{p_k\}_{k=1}^N,~p_k\in(2\ZZ-2k-1)/4 \atop p_{k+1}-p_k\in
  \ZZ_{\geq 0}+1/2} |q|^{D(\{p_k\})+\sum_{k=1}^N |p_k|}
  \left({q\o \bar{q}}\right)^{\sum_{k=1}^N {p_k\o 2}},}
where
\eqn\Dpk{ D(\{p_k\}) = \cases{
  0~~~ & if~~$p_k\in \ZZ-{k\o 2}-{1\o 4}$~~~and ~~$\#\{p_k>0\}$
    ~is~even~~$\leftrightarrow~[\phi_{1,2}]$ \cr
  {1\o 10}~~~ & if~~$p_k\in \ZZ-{k\o 2}-{1\o 4}$~~~and ~~$\#\{p_k>0\}$
    ~is~odd~~~$\leftrightarrow~[\phi_{1,4}]$ \cr
  {1\o 10}~~~ & if~~$p_k\in \ZZ-{k\o 2}+{1\o 4}$~~~and ~~$\#\{p_k>0\}$
    ~is~even~~$\leftrightarrow~[\phi_{1,1}]$ \cr
  0~~~ & if~~$p_k\in \ZZ-{k\o 2}+{1\o 4}$~~~and ~~$\#\{p_k>0\}$
    ~is~odd~~~$\leftrightarrow~[\phi_{1,3}]$ .\cr}}
[Compare this result with
\eqn\ZIsnl{ \eqalign{ Z_{{\cal M}(3,4)} ~=~ & |q|^{-1/24} ~
  \sum_{{N=0\atop N~{\rm even}}}^\infty
  ~\sum_{ \{n_k\}_{k=1}^N \subset \ZZ/2 \atop
  n_{k+1}-n_k \in \ZZ_{\geq 1}} \cr
 &\times ~|q|^{\tilde{D}(\{n_k\})+\sum_{k=1}^N |n_k|}
 \left( {q\o \bar{q}} \right)^{\sum_{k=1}^N n_k/2}~,\cr} }
where ~$\tilde{D}(\{n_k\})={1\o 8}$~ if ~$n_k\in\ZZ$~ and 0 otherwise,
which gives the massive-conformal dictionary for the
low-temperature phase of the Ising field theory.]

Eq.~\Zthfi\ represents the partition function of the CFT as a sum
over even-$N$-``particle'' states. To complete the translation to
the (UV limit of the) massive $N$-kink states
we furthermore need a 1--1 map between the
sets $\{p_k\}_{k=1}^N$ in \Zthfi\ and the sets
{}~$\{n_k\}_{k=1}^{N}\subset {1\o 2}\ZZ$~ with
{}~$n_{k+1}-n_k \in \ZZ_{\geq 1}$. The experience of sect.~3 suggests
using \UVlim\ with ~$\tilde{\del}=\tilde{\del}(\infty)$, which is
equal to $\pi/2$ in our model. Indeed,
\eqn\pknk{ p_k ~=~ n_k+{N+1\o 4}-{k\o 2}~~~~~~~~~~~~~~
   (k=1,2,\ldots,N~{\rm even}) }
implements such a map of the allowed $\{n_k\}_{k=1}^N$ onto
$\{p_k\}_{k=1}^N$ with ~$p_k \in {1\o 2}\ZZ-{k\o 2}-{1\o 4}$~
and ~$p_{k+1}-p_k \in \ZZ_{\geq 0}+{1\o 2}$, which are the sets
summed over in \Zthfi.

Hence we conjecture that the massive multi-kink state labeled by
$\{n_k\}_{k=1}^{N}$ in the `$-$'-phase
of the $\phi_{1,3}$-perturbed ${\cal M}(3,5)$ model comes from a
conformal state of scaling dimension and spin
\eqn\dsthfi{ (d,~s) ~=~
 \left(D(\{p_k\}) +\sum_{k=1}^N |p_k|,~\sum_{k=1}^N n_k \right)~~,}
where the $\{p_k\}$ and $D(\{p_k\})$ are given by eqs.~\pknk\ and
\Dpk. (The map between the $\{n_k\}_{k=1}^{N}$ and the
quasiparticle labels by $\{\bar{I}_j\}_{j=1}^{\bar{m}} \cup
\{I_j\}_{j=1}^m$, where $\bar{m}+m=N$, can be also  obtained,
as for the perturbed ${\cal M}(2,5)$ model.) This correspondence
is consistent with the results of~\rLasMar~for the six lowest
2-kink states in the zero-momentum sector, as well as with
the observation that the spinless conformal states of scaling dimension
$1{9\o 10}, 3{9\o 10}$ and 4  evolve into 4-kink states,
which can be deduced from the plots given there.
(According to \Dpk, the corresponding $\{n_k\}$ are
$\{-{3\o 2},-{1\o 2},{1\o 2},{3\o 2}\}$,
$\{-{5\o 2},-{1\o 2},{1\o 2},{5\o 2}\}$, and
$\{-2,-1,1,2\}$.)

\bigskip
Before closing this section, let us make an amusing observation
about two more integrable single-particle theories which
however cannot be formulated
as perturbations of rational CFTs.
These are the real-coupling affine Toda field theories based
on the affine Lie algebras $A_{1}^{(1)}$ and $A_{2}^{(2)}$,
whose UV limit is rather singular, being described by
a free massless uncompactified scalar field~\rKMtba.
The shifted phase-shifts, as determined from the factorizable
$S$-matrices~\rAFZ~of the two theories, satisfy
{}~$\tilde{\del}(\infty)=\pi$~ independently of the coupling.
Using this as the value of $\tilde{\del}$ in eq.~\UVlim, which
has proven to be useful in the models studied earlier, we
obtain ~$p_k=n_k+{N+1\o 2}-k$~ for the ``single-particle
momenta'' in the UV limit. This relation maps the allowed
massive labels ~$\{n_k\}_{k=1}^N \in \ZZ+{N+1\o 2}$~ with
{}~$n_{k+1}\geq n_k+ 1$~ onto ~$\{p_k\}_{k=1}^N\in\ZZ$~ with
{}~$p_{k+1}\geq p_k$, which is the quantization condition
appropriate for free bosons in a box with periodic boundary
conditions.

\newsec{Discussion}

%\medskip  {\it 5.}~
The eigenstates of the hamiltonian of an integrable, massive,
perturbed rational CFT in finite volume can be characterized
in two alternative ways. One is adequate for the large-volume (IR)
regime, where states are labeled (schematically)
by the Bethe Ansatz quantum numbers   $\{ n_k \}$.
The other is taylored for the CFT (UV) limit, and employs the
massless fermionic quasiparticle labels
$\bigl([\phi]~|~\{ \bar{I}_j\} \cup  \{I_j\} \bigr)$,
where $[\phi]$
specifies the conformal family, or a more algebraic description in
terms of generators of the chiral algebra of the CFT acting on
highest-weight (primary) states; either way,
the conformal label directly gives the conformal dimensions
(equivalently scaling dimension and spin) of the field creating
the state in the unperturbed CFT.

Lacking the tools for computing the full exact
spectrum at all volume,
it seems more viable and still very interesting to find the
correspondence between the two alternative characterizations.
In this paper we conjectured the explicit dictionary between
the IR and UV labels in two simple -- yet nontrivial -- theories.
The simplicity of these
two theories lies in the fact that there is a single (quasi)particle in
their spectrum.
Although generalizations of our work
to  models with more particles
and bigger internal symmetry do not look quite
straightforward, we still think
that some useful and general insight has been gained.

The basic feature of the IR description of the spectrum is that the
quantum numbers $n_k$ are those of (GSO-projected) free fermions.
The interaction
in this description shows itself in the $S$-matrix, which is not
constant as a function of rapidity if the theory is nontrivial
(from the point of view of the spectrum, at least). The single-particle
momenta $p_k$ are then shifted at order ${\cal O}(1/L^2)$
from their free-quantized values $2\pi n_k/L$, but the dispersion
relation remains ~$E(p)=\sqrt{p^2+m^2}$ ~as in a free theory.
(It is important to remember, however, that the total energy of a state
is equal to the sum of the single-particle energies only up to
off-shell exponential corrections.)
 This picture conforms with the usual
(perturbative) quantum field theoretical viewpoint of trivial statistics
and short-range interactions which are probed in scattering processes.

The CFT description, on the other hand, is manifestly non-perturbative.
Nontriviality of a theory is indicated already by the presence of
non-half-integral conformal dimensions. It is further reflected in
the quasiparticle picture by nontrivial restrictions on the
allowed quasiparticle labels ${I}_j$; these restrictions are collective
in nature, being dependent on the total number of quasiparticles in
a state. This feature, combined with the fact
that within each conformal family
all levels are equally spaced and the total scaled energy is
{\it exactly} the sum of the single-quasiparticle
ones (which are linear in the
$I_j$), suggests an interpretation of
a nontrivial  CFT as describing free massless ``particles''
obeying ``generalized statistics''.

Hence a massive-conformal dictionary reconciles and provides a bridge
between the two  pictures.
In particular, such a dictionary
can be used to represent the CFT partition
function as a sum over multi-particle states with the
momenta of the ``constituent particles'' being quantized like
ordinary free fermions,
 but with a nontrivial dispersion relation (cf.~\dict\ and \dsthfi).
This observation may provide a clue for generalizations to
other models. The next simplest class of models, after the
ones considered here,
consists of theories with diagonal $S$-matrix
but more than one type of particle.
The strategy we propose for obtaining a (conjecture for a)
massive-conformal dictionary in such theories
is summarized by the following vague prescription:
find a ``nice'' representation of the partition function
of the rational UV CFT as a sum over multi-particle states
labeled by quantum numbers of several types of
free fermions.

In the single-particle cases we studied, such
``nice'' representations -- whose validity is independent of the
question of the massive-conformal
correspondence -- were obtained in two steps.
The first involved the
recasting of the CFT partition function, as expressed via
fermionic quasiparticle representations for the characters,
in a form in which the separate sums over
left- and right-movers are combined into a single sum.
The states summed over in this single sum, which are
still restricted by some ``generalized fermionic statistics'',
were then mapped in the second step onto the required
ordinary fermionic states, using inspiration from
the Bethe Ansatz description of the large-volume levels
in the perturbed theory.

As an example of how the first step can be implemented
consider the case of the minimal models ${\cal M}(2,2n+3)$.
Using  Gordon's theorem~\rGord, whose analytical version~\rAnd~is
encountered~\rFNO~in the fermionic $n$-quasiparticle representations
for the relevant Virasoro characters, the following
generalization of eq.~\ZYLp\ is obtained:
\eqn\ZGE{ \eqalign{ Z_{{\cal M}(2,2n+3)}  ~=~ &
  |q|^{-\tilde{c}^{(2,2n+3)}/12}
  ~ \sum_{N=0}^\infty ~\sum_{ \{p_k\}_{k=1}^N \subset \ZZ,
  ~p_{k+1}\geq p_k \atop p_{k+n}-p_k \geq 2} \cr
 &\times ~ |q|^{\hat{d}^{(2,2n+3)}_{1,n+1-\#\{p_k=0\}}
   + \sum_{k=1}^N |p_k|}
  \left( {q\o \bar{q}} \right)^{\sum_{k=1}^N {p_k/2}}~,\cr} }
where ~$\tilde{c}^{(2,2n+3)}={2n\o 2n+3}$~ and
{}~$\hat{d}^{(2,2n+3)}_{1,r}={n(n+1)-(r-1)(2n+2-r)\o 2n+3}$
{}~($r=1,2,\ldots,n+1$).
The restriction of having no more than $n$ ``momenta'' $p_k$
of the same value, imposed in \ZGE, suggests that we are
dealing with $n$ different types of particles, which is
indeed the case in both the integrable $\phi_{1,3}$- and
$\phi_{1,2}$-perturbations of ${\cal M}(2,2n+3)$~\rFKMS\rSKMM.
However, in \ZGE\ these $n$ types of particles appear
``indistinguishable'', which unfortunately
prevents a straightforward use
of this formula for implementing the second step mentioned
above. Hence disentangling the massive-conformal dictionary
in these and other theories remains an intriguing
challenge.

\medskip
\bigskip
{\it Acknowledgements.}
I wish to thank T.R.~Klassen and B.M.~McCoy for discussions.
This work was supported in part by the US-Israel Binational
Science Foundation.

\vfill\eject
\listrefs

\bye\end